# Determination of the Riemann modulus and sheet resistance of a sample with a hole by the van der Pauw method

K. Szymański, K. Łapiński and J. L. Cieśliński

*University of Białystok, Faculty of Physics, K. Ciołkowskiego Street 1L, 15-245 Białystok, Poland*

The van der Pauw method for two-dimensional samples of arbitrary shape with an isolated hole is considered. Correlations between extreme values of the resistances allow one to determine the specific resistivity of the sample and the dimensionless parameter related to the geometry of the isolated hole, known as the Riemann modulus. The parameter is invariant under conformal mappings. Experimental verification of the method is presented.

**I. Introduction**

Determination of direct relations between geometrical size or shape of the sample and electrical quantities is of particular importance as new measurement techniques may appear or the experimental accuracy can be increased. As an example, we indicate a particular class of capacitors in which cross-capacitance per unit length is independent of their cross-sectional dimension [1]. Development of this class of capacitors leads to a new realization of the unit of capacitance and the unit of resistance [2-4]. The second example is the famous van der Pauw method of measuring the resistivity of a two-dimensional, isotropic sample with four contacts located at its edge [5,6]. The power of the method lies in its ability to measure the resistances with four contacts located at arbitrarily located point on the edge of a uniform sample. The result of the measurement is sheet resistance, i.e., the ratio of specific resistivity and thickness. Both methods are formally closely related and can be considered as dual [7].

The van der Pauw method is widely used in laboratory measurement techniques. However, strict assumptions of the method – point contacts and the sample homogeneity – are difficult to obey in experimental practice. The resistivity measurement is a weighted averaging of local resistivities. Koon *et al* has introduced concept of a weighting function used for estimation of the resistance as a weighted averaging of local resistivities. A formalism to calculate the weighting function for van der Pauw samples was proposed and measured in [8]. However the weighting function has to be calculated for each sample shape [9] making its practical use inconvenient. Many authors have demonstrated that sample inhomogeneity or defects results in incorrect value of the measured resistivity and proposed some correcting procedures [9,10,11,12]. To our best knowledge all of the proposed corrections depend on the specific sample shape and thus spoil the generality of the van der Pauw approach.

Investigations of the relationship between the sample geometry, some specific probe positions and the precision of the van der Pauw type measurements were presented in [13]. An improved numerical approach to the classical van der Pauw method was given in [14]. Quantitative influence of defects on measurements of resistivity of thin films by van der Pauw method was investigated in [11], while the uncertainty of the surface resistivity in some textiles showing electric conductivity electric current was estimated in [15]. It was experimentally verified that presence of a hole negatively influences van der Pauw measurements [11], but until our recent paper [16] no attempts were made to generalize and extend the van der Pauw method on samples with holes.

In this paper we extend four-point van der Pauw technique on doubly-connected samples (i.e., samples with a single hole). We keep all other assumptions associated with the van der Pauw method (homogeneous infinitesimally thin sample, sufficiently small contacts etc.). We used two axially symmetric shapes of such samples: an annulus and a thin-walled cylinder as well as an irregular shape with an isolated hole. All cases are mathematically equivalent, see [17]. Any fixed choice of four probes yields a pair of resistances. We recall that in the case of samples without holes these resistances, divided by specific resistivity of the sample and multiplied by its thickness, form dimensionless pairs, say ($x$,$y$), which lie on a single curve given by the famous van der Pauw equation $\exp(-\pi x)+\exp(-\pi y)=1$. Recently, explicit results of the van der Pauw method for a sample containing an isolated hole were presented [16]. Results of measurements and numerical analysis strongly suggested that in the case of doubly-connected samples the four-probe resistances obey an inequality, $\exp(-\pi x)+\exp(-\pi y)\leq 1$, similar in the form to the van der Pauw equation. We propose here a second boundary curve (or an envelope). The shape of the envelope depends on a single dimensionless parameter (known as modulus or the Riemann modulus, see [17]) characterizing the doubly-connected sample. We conjecture that for any given sample with a hole all four-probe resistances lie in a region between these two curves. This conjecture, supported by many experimental and numerical tests, is an interesting mathematical open problem.

Another result of this paper is an experimental procedure for determining the envelope. As a consequence we obtain two important quantitative results: performing a series of measurements by changing location of the four probes, we are able to determine both Riemann modulus and sheet resistance of any doubly-connected homogeneous sample.

## 2. Four-point method on a sample with an isolated hole

Let us consider a two dimensional conductive medium with four different point contacts located at points $\alpha, \beta, \gamma, \delta$ on the sample edge. A current $j_{\alpha\beta}$ enters the sample at the contact $\alpha$ and leaves at the contact $\beta$, while the potential $V_{\gamma\delta}$ is measured between contacts $\gamma$ and $\delta$. The four-probe resistance for contacts $\alpha, \beta, \gamma, \delta$ is defined as $R_{\alpha\beta\gamma\delta}=V_{\gamma\delta}/j_{\alpha\beta}$, see Fig. 1. It was shown [5], that for a sample without a hole (Fig. 1a) two resistances, $R_{\alpha\beta\gamma\delta}$ and $R_{\beta\gamma\delta\alpha}$, fulfill the relation

$$\exp(-R_{\alpha\beta\gamma\delta}/\lambda) + \exp(-R_{\beta\gamma\delta\alpha}/\lambda) = 1, \qquad (1)$$

where $\lambda=\rho/(\pi d)$ and $\rho$ is the specific resistivity of the sample with thickness $d$. Eq (1) can be considered also as a correlation between resistances $R_{\alpha\beta\gamma\delta}$ and $R_{\beta\gamma\delta\alpha}$ when measurements of four-probe resistances are performed many times for different positions of the contacts $\alpha, \beta, \gamma, \delta$. In the simply-connected case (Fig. 1a) the quantities $R_{\alpha\beta\gamma\delta}/\lambda$ and $R_{\beta\gamma\delta\alpha}/\lambda$ are located on the one universal curve given by (1).

The same quantities, $R_{\alpha\beta\gamma\delta}/\lambda$ and $R_{\beta\gamma\delta\alpha}/\lambda$, measured on a doubly-connected sample (sample with a single isolated hole) do not fit to any universal curve and the points ($R_{\alpha\beta\gamma\delta}/\lambda$, $R_{\beta\gamma\delta\alpha}/\lambda$) are dispersed between the universal van der Pauw curve (defined by Eq. (1), solid lines in Figs. 1ab) and another curve (dashed lines in Figs. 1bc) which will be defined below.

It was shown recently [16] that in the case of four contacts located at positions $\alpha, \beta, \gamma, \delta$ ($0 \leq \alpha \leq \beta \leq \gamma \leq \delta \leq 2\pi$) at the same edge of an infinitesimally-thin-walled cylinder of finite height $H$ and radius $r$ (shown in Fig. 2a), the four probe resistance is given by

$$R_{\alpha\beta\gamma\delta} = \lambda \ln \frac{G(\gamma-\alpha, h)G(\delta-\beta, h)}{G(\gamma-\beta, h)G(\delta-\alpha, h)}, \qquad (2)$$

where $h=2H/r$, $\lambda=\rho/(\pi d)$ and $\rho$ is specific resistivity of the cylinder with infinitesimal thickness of the wall $d$, and, finally, the function $G(\phi,h)$ is defined as

$$G(\phi, h) = \left|\sin\frac{\phi}{2}\right| \prod_{n=1}^{\infty}\left(1 - \frac{\cos\phi}{\cosh hn}\right). \qquad (3)$$

The parameter $h$ is related to the parameter $\mu$ which is known as modulus or Riemann modulus of the doubly-connected sample. Any two-dimensional sample with a hole can be transformed by conformal mapping into an annulus (see [18], page 255). In [16] we have shown how to transform the cylinder (Fig. 2a) into an annulus. The ratio of radii of the annulus is called the Riemann modulus $\mu=r_{out}/r_{inn}$ ($\mu>1$), see [17]. Because potentials are invariant under conformal mapping, four-probe resistances of a sample with a hole are equivalent to the four-probe resistances of some annulus. On the other hand, any annulus is conformally equivalent to some thin-walled cylinder, see [16]. Hence there follows the relation $\mu=\exp(h/2)$.

We have proposed [16] that resistances $R_{\alpha\beta\gamma\delta}$ and $R_{\beta\gamma\delta\alpha}$ for a sample with an isolated hole (i.e., resistances given by Eq. (2)) obey the inequality:

$$\exp(-R_{\alpha\beta\gamma\delta}/\lambda) + \exp(-R_{\beta\gamma\delta\alpha}/\lambda) \leq 1. \qquad (4)$$

The rigorous proof of (4) is not known yet. The inequality is supported by numerical evidences and results of many our measurements, as usual within a finite experimental uncertainty. An example of experimental verification is presented in Fig. 3 of [16]. It agrees with earlier van der Pauw measurements [11] performed on the samples with holes. The inequality (4) shows that points ($R_{\alpha\beta\gamma\delta}/\lambda$, $R_{\beta\gamma\delta\alpha}/\lambda$) corresponding to a sample with a hole do not fit to any universal curve. They are located (see Fig. 1b) above the universal van der Pauw curve (solid line) and below the curve (dashed line) given parametrically by the set of two equations

$$\frac{R_{\alpha\beta\gamma\delta}}{\lambda} = 2\ln\frac{G(\pi,h)}{G(\pi-\phi,h)}, \qquad \frac{R_{\beta\gamma\delta\alpha}}{\lambda} = 2\ln\frac{G(\pi,h)}{G(\phi,h)}. \qquad (5)$$

The curve (5), parameterized by $\phi$ ($0\leq\phi\leq\pi$) and shown in Figs. 1bc by the dotted line, will be called the envelope.

The essence of this new result consists in the observation that there exists an extreme curve which bounds values of ($R_{\alpha\beta\gamma\delta}$, $R_{\beta\gamma\delta\alpha}$) from the top (see Fig. 1b). In the case of axially symmetric sample this curve corresponds to $\alpha$, $\beta$, $\gamma$ and $\delta$ located at vertices of a rectangle, i.e. $\delta-\beta=\pi$, $\gamma-\alpha=\pi$. The parametric form of this curve is given by (5). The procedure for its experimental determination, including the sample without axial symmetry, is given in the next section. Our result, confirmed by many numerical and experimental tests but lacking a rigorous mathematical proof, can be also stated purely in terms of the function $G$ defined by (3). On the Cartesian plane ($x$, $y$) we consider the region $\Omega_h$ bounded by the straight line $x+y=1$ and by the curve defined parametrically by

$$x = \frac{G(\pi,h)}{G(\pi-\phi,h)}, \qquad y = \frac{G(\pi,h)}{G(\phi,h)}, \qquad 0\leq\phi\leq\pi. \qquad (6)$$

We define

$$x_{\alpha\beta\gamma\delta} := \frac{G(\gamma-\beta,h)G(\delta-\alpha,h)}{G(\gamma-\alpha,h)G(\delta-\beta,h)}, \qquad y_{\alpha\beta\gamma\delta} := \frac{G(\delta-\gamma,h)G(\alpha-\beta,h)}{G(\delta-\beta,h)G(\alpha-\gamma,h)}. \qquad (7)$$

We conjecture that ($x_{\alpha\beta\gamma\delta}$, $y_{\alpha\beta\gamma\delta}$)$\in\Omega_h$ for any $0\leq\alpha\leq\beta\leq\gamma\leq\delta\leq 2\pi$. The proof of this statement is an open problem. The reader should not be confused by the same abbreviation ($x$, $y$) used in (6) and in the introduction.

The result (5), although deduced for a specific case of axially symmetric sample with a hole (the cylinder), is invariant under any conformal transformation and is valid for a flat sample of any shape with a single isolated hole.

## 3. Experimental determination of the envelope

Since pairs of extreme resistances $R_{\alpha\beta\gamma\delta}$ and $R_{\beta\gamma\delta\alpha}$ obeying (5) are located on a curve (dotted line in Fig.1b) dependent on two parameters $h$ and $\lambda$ (directly related to the Riemann modulus and sheet resistance of the considered sample), it is possible to determine these parameters by performing the standard fit. Therefore, the crucial point is to determine experimentally the envelope given by Eq. (5). The determination of the envelope is easy in the case of axially symmetric samples, like an annulus or a cylinder, because the symmetry of such samples allows for an easy location of contacts at the vertices of the rectangle (inset in Fig.1c). In the case of an arbitrary shape (Fig. 1b) it is not obvious how to arrange the positions of contacts to measure extreme values of the resistances. There is a possibility of measuring at random a large number of resistances $R_{\alpha\beta\gamma\delta}$ and $R_{\beta\gamma\delta\alpha}$ for various positions of contacts. In this way the shaded area shown schematically in Fig. 1b would be filled (in a relatively dense way) by experimental points. However, this procedure seems to be time-consuming and not practical.

We propose a procedure of the envelope determination, valid for a shape with axial symmetry (an infinitesimally-thin-walled cylinder shown in Fig. 2$a$, or an annulus - inset in Fig. 1c), which can be easily extended on samples of any shape. For an arbitrary position of the current contacts $\alpha$, $\gamma$, there is a continuum of equipotential pairs of points $\beta_0$, $\delta_0$ (Fig. 2d). Let us consider four-point resistances $R_{\alpha\beta_0\gamma\delta_0}$. Among the pairs $\beta_0$, $\delta_0$ there is a pair $\beta_{ext}$, $\delta_{ext}$, for which the resistance $R_{\alpha\beta_{ext}\gamma\delta_{ext}}$ has a local extreme. The pair $\beta_{ext}$, $\delta_{ext}$ is located at the diameter of the axially symmetric sample and is located symmetrically with respect to $\alpha$, $\gamma$ (see Fig. 2d). Although the construction of finding the diameter $\beta_{ext}$, $\delta_{ext}$ was performed on an axially symmetric sample, during the construction the symmetry was not needed. The procedure of determining extreme values $\beta_{ext}$ and $\delta_{ext}$ can be performed for a pair of contacts $\alpha$, $\gamma$ located arbitrarily on the edge of a sample of any shape (Fig. 2e).

For another choice of current contacts $\alpha'$, $\gamma'$, a second diameter $\beta'_{ext}$, $\delta'_{ext}$ can be found (see Fig. 1f). The two diameters $\beta_{ext}$, $\delta_{ext}$ and $\beta'_{ext}$, $\delta'_{ext}$ determine vertices of a rectangle. Thus, the resistance measured with current contacts at $\beta'_{ext}$, $\beta_{ext}$, and voltage contacts at $\delta'_{ext}$, $\delta_{ext}$ (or current contacts at $\beta_{ext}$, $\delta'_{ext}$ and voltage contacts at $\delta_{ext}$, $\beta'_{ext}$, $\beta_{ext}$ in Fig. 2e,f) belongs to the envelope (5).

The presented procedure was verified experimentally. Three flat samples were prepared with the same material. Two of them had an axially symmetric shape, disk and annulus, while the third one was irregular in shape. All three samples were measured by the four-probe method and the resistances related to the envelope (5) are shown in Fig. 3. All theoretical lines are drawn with just one common parameter $\lambda$ found from the best fits. The parameter $h=\infty$ was used for the disk while $h=2\ln(R/r)$ for the annulus with the inner radius $r$ and outer radius $R$, matching perfectly the theoretical lines (5) and the experimental points. For the sample with the irregular shape, the parameter $h=1.455$ was found from the best fit (see dashed line in Fig. 3).

## 4. Discussion

We have demonstrated that standard van der Pauw-type measurements can be extended on doubly-connected samples (samples with a hole). We are able to determine the sheet resistance and conformally invariant characteristics of such samples – a dimensionless geometrical parameter directly related to the Riemann modulus. In specific cases parameter $h$ is simply expressed by the sample size: $h=2H/r$ for infinitesimally-thin-walled cylinder of height $H$ and radius $r$, while $h=2\ln(r_{out}/r_{inn})$ for an annulus with inner radius $r_{inn}$ and outer radius $r_{out}$, see [16].

The novelty of the method is the experimental determination of the geometrical parameter (or the Riemann modulus) by the measurements of the extreme values of the four-probe resistances. For a given sample with a hole the extreme resistances are located on the curve given by Eq. (2) which allows to determine two independent sample characteristics – the geometrical parameter $h$ and the ratio of the specific resistivity to the thickness, $\lambda$. The disadvantage of the presented method is the necessity to perform a series of measurements changing positions of the contacts. We hope to overcome these difficulties. The work in this direction is in progress.

The great advantage of the method is its applicability to a sample of any shape. The parameter $h$ has the same value within the class of samples obtained by conformal mapping. In particular, the annulus can be transformed to a shape with a linear slit. Potential applications include detection of linear cracks and detection of a non-conducting inhomogeneity in otherwise conducting sample. The presented method is important for practical applications because the parameter $h$ (or the Riemann modulus) can be interpreted as a measure of the sample inhomogeneity. The apparent advantage of the method is its global character – the hole can be detected irrespectively of its location.

**Figures**

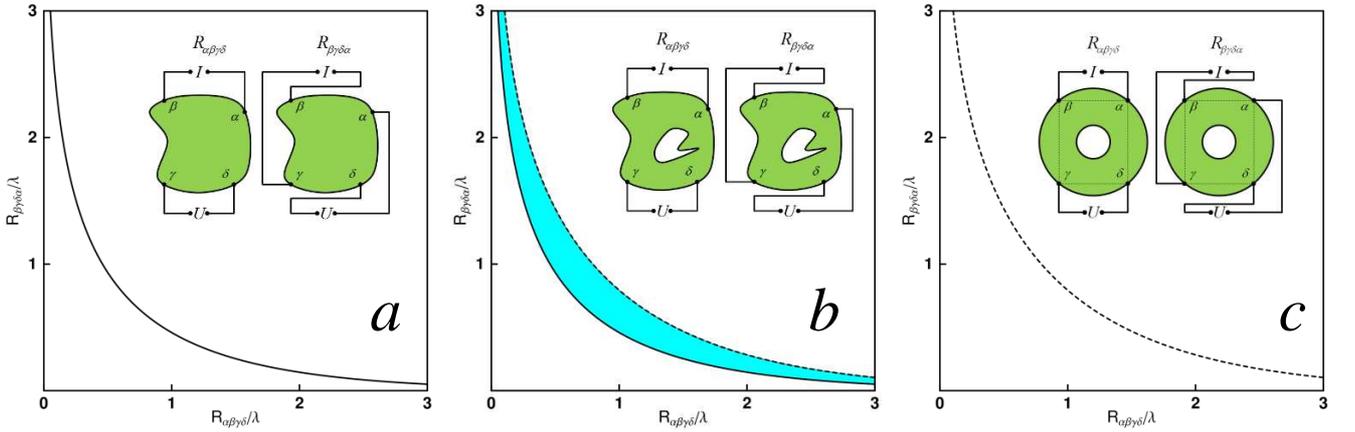

Fig. 1. a) Expected correlation between quantities $R_{\alpha\beta\gamma\delta}/\lambda$ and $R_{\beta\gamma\delta\alpha}/\lambda$, where $\alpha, \beta, \gamma, \delta$ are arbitrary positions of contacts for a sample of any shape without a hole b) shaded area between the solid and the dashed line showing dispersion of quantities $R_{\alpha\beta\gamma\delta}/\lambda$ and $R_{\beta\gamma\delta\alpha}/\lambda$ for sample with a single isolated hole c) correlation between extreme values of quantities $R_{\alpha\beta\gamma\delta}/\lambda$ and $R_{\beta\gamma\delta\alpha}/\lambda$ for a sample with a single isolated hole. For an annulus the extreme values can be achieved by arrangement of contacts at the vertices of a rectangle shown by the dotted lines in the inset. Inset: scheme of measurement of $R_{\alpha\beta\gamma\delta}$ and $R_{\beta\gamma\delta\alpha}$ resistances for the contacts located at positions $\alpha\beta\gamma\delta$.

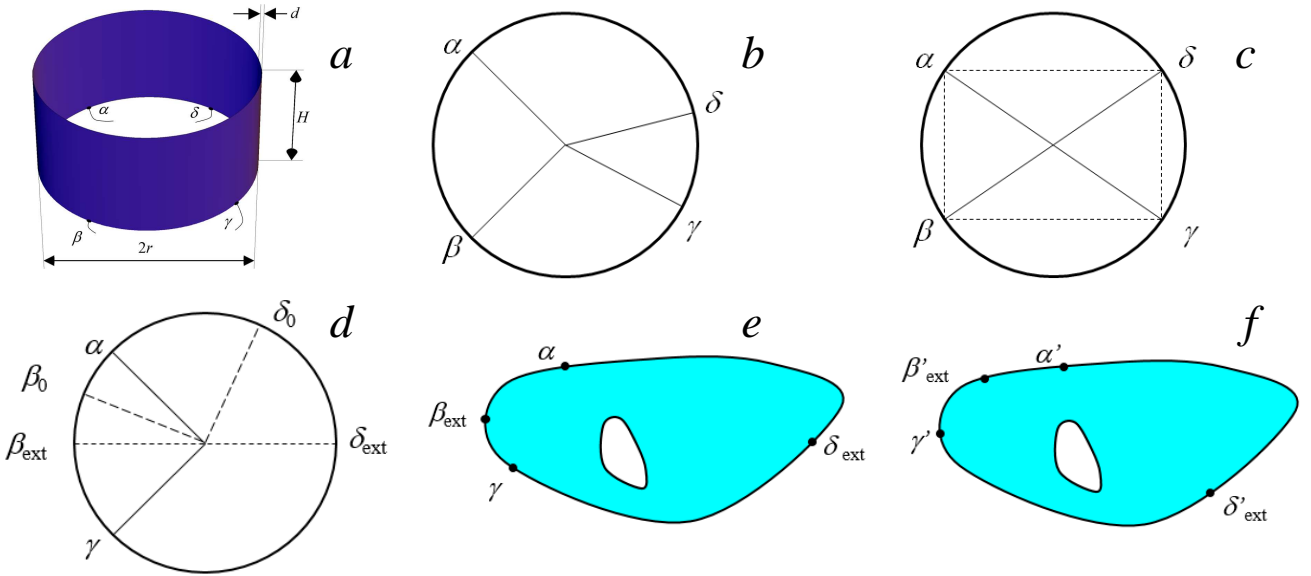

Fig. 2 *a*) 3-D view of an infinitesimally-thin-walled cylinder of finite height $H$, radius $r$ and wall thickness $d$, with four contacts at arbitrary positions $\alpha\ \beta\ \gamma\ \delta$ on the same edge. The cylinder is conformally equivalent to the shapes with holes shown in the inset of Fig. 1. *b*) Schematic view of an axially symmetric sample (cylinder or annulus). *c*) Scheme of the contact positions resulting in an extreme value of the resistances $R_{\alpha\beta\gamma\delta}$ and $R_{\beta\gamma\delta\alpha}$. *d*) Positions of equipotential pair of contacts $\beta_0, \delta_0$ and pair $\beta_{ext}, \delta_{ext}$ for determination of the extreme values of the resistances (see text) for an axially symmetric sample. *e*) Positions of contacts $\beta_{ext}, \delta_{ext}$ for determination of the extreme values of the resistances on a flat sample with a hole. *f*) Another positions of contacts $\beta'_{ext}, \delta'_{ext}$ for determination of the extreme values of the resistances.

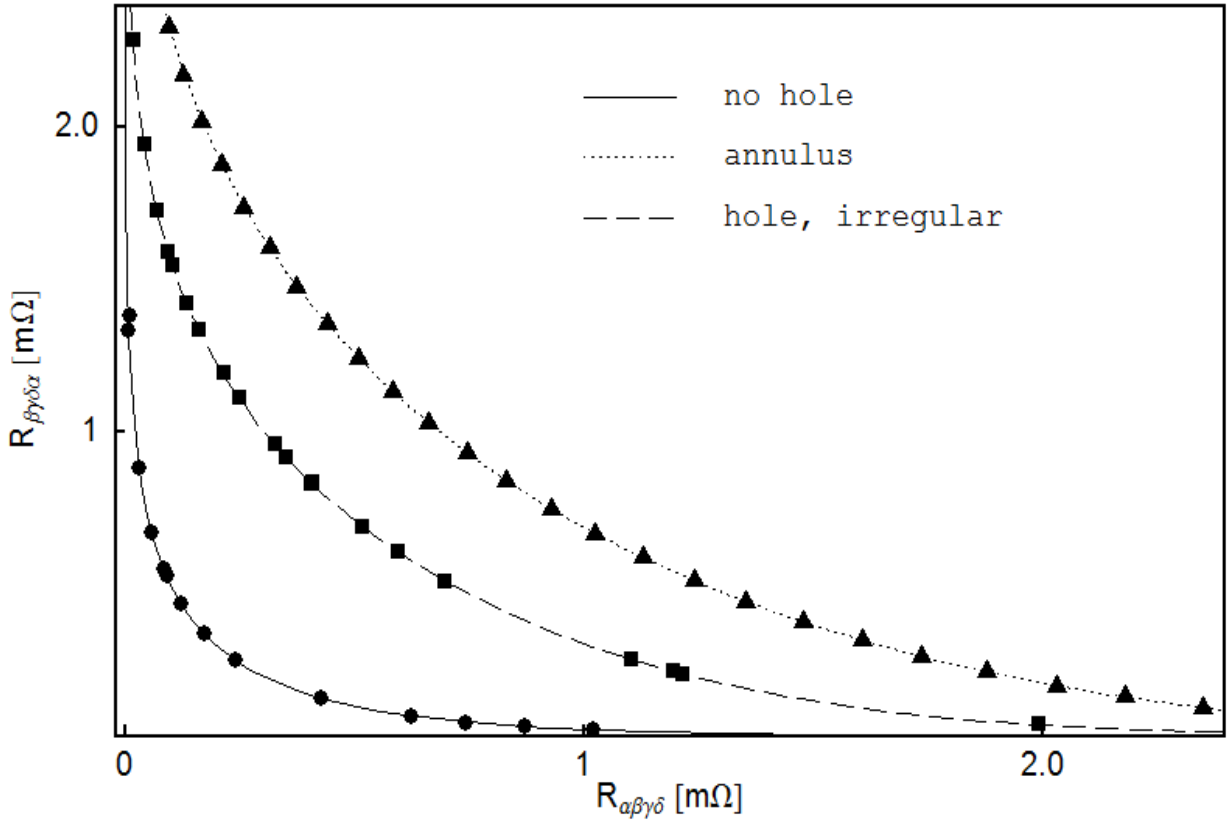

Fig. 3. Correlation between measured resistances $R_{\alpha\beta\gamma\delta}$ and $R_{\beta\gamma\delta\alpha}$ for sample without hole (van der Pauw case, circles), annulus with radiuses $r_{inn}$=74.5 mm and $r_{out}$=125.5 mm (triangles), and sample of irregular shape with an isolated hole (squares). All lines correspond to the same value of the ratio of the specific resistivity and the sample thickness, $\lambda$. The lines correspond to eq. (2) with $h=\infty$ (van der Pauw case, solid line), $h=2\ln(r_{out}/r_{inn})$ (dotted line), and $h=1.455$ found from the best fit (dashed line).

**References**

[1] A. M. Thompson and D. G. Lampard, Nature 177, 888 (1956).

[2] W. K. Clothier, Metrologia 1, 36 (1965).

[3] A. M. Thompson, Metrologia 4, 1 (1968).

[4] H. Bachmair, The European Physical Journal Special Topics 172, 257 (2009).

[5] L. J. van der Pauw, Philips Research Reports 13, 1 (1958).

[6] L. J. van der Pauw, Philips Technical Review 20, 220 (1958).

[7] J. J. Mareš, P. Hubík and J. Krištofik, Meas. Sci. Technol. 23, 045004 (2012).

[8] D. W. Koon and C. J. Knickerbocker, Rev. Sci. Instrum. 63, 207 (1992).

[9] D. W. Koon, F. Wang, D. H. Petersen, and O. Hansen, J. of Appl. Phys. 114, 163710 (2013).

[10] O. Bierwagen, T. Ive, C. G. Van de Walle and J. S. Speck, Appl. Phys. Lett. 93, 242108 (2008).

[11] J. Náhlík, I. Kašpárková and P. Fitl, Measurement 44,1968 (2011).

[12] F. Wang, D. H. Petersen, M. Hansen, T. R. Henriksen, P. Bøggild, and O. Hansen, J. Vac. Technol. B 28, C1 (2010).

[13] S. Thorsteinsson, F. Wang, D. H. Petersen, T. M. Hansen, D. Kjær, R. Lin, J-Y. Kim, P. F. Nielsen, and O. Hansen, Rev. Sci. Instrum 80, 053902 (2009).

[14] J. L. Cieśliński, Thin Solid Films 522, 314 (2012).

[15] M. Tokarska, IEEE Transactions on Instrumentation and Measurement 63, 1593 (2014).

[16] K. Szymański, J. L. Cieśliński and K. Łapiński, Phys. Lett. A 377, 651 (2013).

[17] Z. Nehari, Conformal mapping, McGraw-Hill, New York 1952.

[18] L. V. Ahlfors, Complex analysis, Third edition, Mc Graw-Hill, 1979.